# Present and future frameworks of theoretical neuroscience: outcomes of a community discussion

Horacio G. Rotstein[1] and Fidel Santamaria[2]

1 New Jersey Institute of Technology, Newark, NJ 07102  horacio@njit.edu

2 The University of Texas at San Antonio, San Antonio, TX 78249
fidel.santamaria@utsa.edu

## Abstract

We organized a workshop on the 'Present and Future Frameworks of Theoretical Neuroscience', with the support of the National Science Foundation. The objective was to identify the challenges and strategies that this field will need to tackle in order to incorporate vast and multi-scale streams of experimental data from the technologies developed by the BRAIN initiative. The participants, divided in workgroups, identified five key areas that, while not exhaustive, cover multiple aspects of current challenges needed to be developed: Dynamics-statistics; multi-scale integration; coding; brain-body integration; and structure of neuroscience theories. While each area is different, there were coincidences on finding theoretical paths to incorporate biophysics, energetics, and ethology with more abstract coding and computational approaches. Each workgroup has continued to work   after the meeting to develop the ideas seeded there, which are started to being published. Here, we provide a perspective of the discussions of each workgroup that point to building on the present foundations of theoretical neuroscience and extend them by incorporating multi-scale information with the objective of providing mechanistic insights into the nervous system.

## Introduction

Theoretical neuroscience has provided insights about the functioning of the nervous systems at all physical scales (1-7). From ion channel activation, action potential generation, and synaptic connectivity, to network dynamics, navigation, learning & memory, decision making, and behavior. However, integrating different theoretical methodologies in the process of generation of knowledge and transforming theories and models from one scale, system, function, and species to another are current challenges of any theoretical or computational approach that aims to understand the nervous system. These challenges are more pressing as theories should integrate the expected vast data streams derived from the use of new technologies product of the BRAIN initiative (8).

We identify three primary practical objectives of theoretical neuroscience: to provide quantitative testable predictions of brain function, to explain neuronal mechanisms in healthy and diseased brains, and to abstract computational strategies to develop better engineering tools. Achieving these goals requires developing theoretical approaches that encompass the shared assumptions of models developed for different species and scales. If this is in fact possible, then experimental neuroscience would benefit by having access to theoretical frameworks that can take data points obtained at one scale of organization and species and apply them to other scales and species. Therefore, it is necessary to examine the present existing frameworks and determine whether they point to one or several theoretical approaches, what new frameworks need to be developed and how to link them. It is no less important to clarify what is a theoretical framework and how frameworks are different from theories and models, and how different modeling approaches can be integrated to produce a cohesive body of knowledge. These frameworks should build on the existing conceptual structure of the field and should not be compartmentalized, but must span across scales, e.g. spatial, temporal; levels of organization, e.g. subcellular, cellular, microcircuit, complex networks; systems; species; and theoretical sub-disciplines.

With these ideas in mind, we organized an NSF-supported workshop, "Present and Future Frameworks on Theoretical Neuroscience", which we conceived as a discussion forum rather than a series of talks about current research. We brought together several dozen experimental, computational, and theoretical researchers to discuss the structure of the state-of-the-art, and then identify fields within neuroscience that could benefit from developing frameworks that cross scales and established disciplines. The objective of the workshop was to generate ideas that could be transformed into distributable products that can serve as a guide for program officers in public and private organizations, researchers, and the public. We envision these products to provide structure and to engage the community into a formal discussion on how to advance theoretical neuroscience.

## The straw man strategy

Since the objective of the workshop was to engage the participants in discussions about general frameworks of theoretical neuroscience and not to focus on their individual projects, it was necessary to come up with a different approach to traditional workshops to generate productive conversations. We designed a number of activities that included pre-meeting work focused on a white paper, the participation of a philosopher of science to help understand the structure of scientific theories in biology, and round table discussions that integrated the ideas of talented PhD students/postdocs with the working groups. Graduate students and postdocs also had pre-meeting activities where they came up with their own ideas on several proposed topics and, divided in groups and mentored by faculty participants, prepared posters on these ideas that were actively discussed during the meeting.

The first version of the pre-meeting white paper was available on-line to the participants and was deliberately designed to generate and organize the discussion. In this document we built a straw man argument, a proposal that we knew would not be of general agreement, and that might be irritating to some, but that would be helpful as a starting point. We claimed that theoretical neuroscience is a collection of ideas and tools that depend on various factors, e.g. the scale of the system, and vary according to the applications rather than a visible coherent picture. As an *a priori* organization we proposed to divide theoretical neuroscience into axes: Information-Biophysics, Precise time-Statistical coding, Network modularity-Scale free, Dynamics-Statistics. These axes do not represent a dichotomy but a dimension along which theories and models are built. We then argued that whatever the shape or structure of these axes might be, they should be applied on the broader objectives of the field: Integration across scales, from synapses to systems; and across species, from invertebrates to vertebrates. In parallel, we proposed that whatever coding strategy is developed it must be bounded by evolutionary principles that constraint efficiency and energetic costs.

## The future frameworks

The objective of the first day of the meeting was to discuss the ideas based on the white paper and organize working groups. However, we determined that it was of great importance to differentiate between collections of models and theory. In order to create a productive dialogue, we invited a philosopher of science, Dr. William Bechtel who presented his work on the structure of biological theories (9, 10). The subsequent discussion based on his talk helped then divide the participants in groups with the task to describe the state-of-the-art in their subfields and then propose a road map for future framework. We summarize below the workgroups and their conclusions at the end of the workshop.

**Structure of neuroscience theories workgroup**

Perhaps neurobiology is the first biological discipline where theoretical and computational approaches have been intricately used with experiments to explain mechanisms and generate predictions that have been successfully tested experimentally.  However, there is still confusion about the structure of theories and how they interact with experimental science, and the role of models within theories. Furthermore, the concept of "framework", which is central to our endeavor, and how frameworks differentiate from theories and models require a deep examination, in the context of neuroscience research. The advent of new technologies that will allow interrogating the nervous system at multiple levels makes it important to develop new theoretical frameworks to integrate the exponentially growing data streams with the goal to determine mechanistic or functional relationship of the underlying processes. The challenge of the present framework is to find an effective structure that allows to categorize experiments and computer simulations within theoretical frameworks.

This workgroup took the pragmatic view that "scientific practice is a problem-solving endeavor", differently from the traditional theory identification and falsification process. Instead of proposing an overarching theory in neuroscience, they proposed to organize knowledge in terms of frameworks, theories, and models, and to use three categories for each of them: descriptive, mechanistic, and normative. These frameworks, theories and models are context-dependent, and do not belong to any specific category in the absence of a description of the empirical problem they intend to solve. In brief, the descriptive category consists of phenomenological approaches to neuronal processes; mechanistic theories aim to describe the interactions of the nervous system based on underlying components; and normative theories use the biological function of a system under the constraints of an objective function. This workgroup has recently published a thorough development of these ideas (11).

**Dynamics-Statistics workgroup**

At large, the dynamical and statistical neuroscience communities have approached problem-solving and the generation of theories and models from different perspectives, using different tools, a different language and with different mindsets (12). While there is some overlapping between the different approaches (e.g., the use of stochastic differential equations combines ideas from both fields as well as certain parameter estimation tools), a conceptual separation between the way the two fields address theoretical neuroscience questions remains prominent. However, there is no *a priori* reason for neuroscience frameworks, theories and models to be described primarily in terms of one or the other, and theoretical neuroscience will certainly profit from combined approaches.

This group addressed these issues and focused on expanding the understanding of the current state of the art, a path initiated in (12), and on how the different fields can collectively contribute to the generation of common frameworks, theories and models to solve the empirical problems that emerge from the existing and newly collected data. In other words, how to develop a common language and combine statistical and dynamic tools to provide appropriate descriptions and mechanistic explanations of these data. For example, mechanistic studies using dynamical models are typically carried out by using dynamical systems tools and ideas or numerical simulations, while statistical neuroscience will use the theory of causality, which, on the other hand, is engrained in the idea of a dynamical system. Are these approaches different? Or are different faces of the same coin? How to formulate problems that can benefit from both points of view?

**Multi-scale theories workgroup**

The nervous system is divided by biological functions which could integrate multi-scale processes, for example the detection of a single photon or sensory transduction of sound. Presently, instead, most models and theories focus on a physical or temporal scale. In many cases the phenomena outside the scale of interest is averaged out in the form of a mean field response or is considered noise. The challenge is to develop frameworks and theories that allow the building of models at one scale to be functionally integrated into other scales. This will be more important as multi-scale information of identified behavior is integrated from the molecular to the electrophysiological and behavioral levels.

This workgroup started from a discussion on the present large-scale efforts to build models of single cells or networks and then integrate those models into larger scale systems. They identified that the assumption of mean-field only applies when the system can be homogenized, and this assumption breaks down when a sub-system consists of multiple heterogeneous components. This workgroup also identified two additional frontiers that are poorly represented in current neural models (of any scale) but may inform theories of neural function across levels: levels of heterogeneity and energetic-metabolic constraints. These concepts are not novel, but our understanding of either remains quite nascent. They highlighted a need for future endeavors to systematically document electrical heterogeneity and energetic constraints-energy metabolism, as well as a need for new mathematical techniques to translate these features between levels of description.

**Coding workgroup**

There was an agreement that there is no single definition of coding. As such, it was identified that there is a need for clarity of the term and how it is applied to different conditions. An interesting discussion was that coding can be understood in two different ways. From a correlative point of view coding, either by spike timing or firing rate strategies is a static correlative process that can be characterized by mutual information. A second form of coding is by the biophysical decoding properties of the downstream receivers. In this case, the same spiking train generated by one neuron could carry different information to other cells. The challenge is to harmonize the definitions of coding to build theories that can accommodate the influx of large-scale experimental data in freely behaving organisms.

This workgroup suggested that it was very important to describe and define the different properties of coding, such as what is meant by firing rate. This could be described as a probabilistic process if looking at a single cell, or the average response of an ensemble. There must be a clear use of time and how a temporal scale maps to one coding strategy or another. Putting all these concepts together, the workgroup has been working on a manuscript in which they detailed the false dichotomy between spike timing vs firing rate codes.

**Brain-body control workgroup**

Currently, theories are based on experimental paradigms that sample the responses of the nervous system, from synapses to behavior, using an impulse-response protocol. One or multiple stimuli are delivered to neurons or organisms using an uncorrelated trial-to-trial approach. This paradigm has provided great advantages in reducing variability and has helped to build causality and mechanistic models of cellular and systems responses. However, these models do not necessarily allow the development of theories that consider the natural environment of organisms. Ultimately, a theory should be able to take data points collected under control conditions to make predictions about the responses of organisms as they interact in the natural world.

This workgroup suggested that in order to build theories of the brain function that are connected to the function of the body it is necessary to design theories, models, and experiments that integrate natural animal behavior. Some examples are closed-loop multi-trial tests where the animal is available to choose the next stimulus, or the presentation depends on past choices. Another strategy is to perform experiments in social contexts, where animals could cooperate or compete, or choose between the two strategies. It was noted that these ideas have been used in other areas, such as ecology and economics, as such, there is a rich repertoire of theories that could benefit neuroscience.

## Training the new generation and engaging society

If we are going to be effective in developing new neuroscience theories, it is important to integrate the new generation of researchers in this task. For this purpose, we requested applications from graduate students and postdocs. Since we wanted these participants to contribute to the conversations, as opposed to just attending passively the workshop, we engaged them before the meeting to select a topic from the themes in the white paper. We asked them to produce a poster working in groups and had them contact faculty participants to mentor and help them in their online discussions.

To be an effective scientist, it is increasingly necessary to improve communications skills. This is particularly important in neuroscience because it is among the research fields that are prone to misinformation and inaccurate reporting (13). Furthermore, theoretical and computational neuroscience increases the difficulty in communicating with the public by being a transdisciplinary field. We provided an opportunity to all participants to engage with the public by partnering with Taste of Science San Antonio (14). Taste of Science is a national non-profit organization that aims to increase science literacy in the adult population by organizing meetings in a casual setting. During the event, about three dozen members of the public had the opportunity to meet scientists not only from the US, but from Europe, Canada, and Japan, and to engage in interesting discussions

While the purpose of the white paper was to prepare the participants to engage in productive conversations during the meeting, we wanted to take a snapshot of the conversations during the meeting. With the help of the UTSA Neurosciences Institute we produced two podcasts. In each podcast we selected from the pool of trainees to chat with faculty. The first podcast took place at the end of the first day. The purpose of this first podcast was to record the impression at the beginning of the meeting on the present state of the field (15). The second podcast was done at the end of the meeting with the objective to report and discuss and compare and contrast the discussion of the different groups on the future frameworks of theoretical neuroscience (16).

## Conclusions

As with any other endeavor that tries to envision how to develop a field the conclusion is better phrased by questions:

1. What are the frameworks to build neuroscience theories? Should we aim to a single framework or multiple, integrated ones? How do they relate to theories and models?
2. Do we have to assume that physical scales determine the biological frameworks? Or, do we have to think first about the biological function to derive the appropriate scale of the problem?
3. How to build neuroscience theories within the broader conceptual framework of Evolution?

4. While theories can be developed for a specific scale, how does one incorporate the heterogeneous effect of the scales not being taken into account?
5. How does one bound theories? Should we use energetic, functional, ethological, or engineering constraints? How do those constraints limit or expand the domain of your theory?
6. While there might be a debate on the statistical or dynamical nature of coding and function of the nervous system, should a theory have mechanisms to bridge between the two strategies? What insight could one derive from such 'interpolation of theories'?

In this article our objective was to report on the process of analysis and initial workshop discussions. Each workgroup has continued working on their ideas, elaborating on multiple issues, raising new ones, and sometimes modifying them. Some of the discussions are becoming manuscripts that will be published in the following months. We hope these papers stir more conversation in the field and attract the interest of theoreticians, experimentalists, and funding agencies.

We would like to emphasize that the goal of the workshop discussions and the papers produced as the results of these discussions is not to provide definite answers or ideas, but to open what we consider is a much needed discussion for the field to move forward and profit from the neuro technological advances in the recent past and these expected in the coming years. Therefore, we welcome new discussions and papers to this series and look forward to additional meetings and discussions within the neuroscience community.

## Acknowledgments

This paper is the result of discussions as part of the workshop "Theoretical and Future Theoretical Frameworks in Neuroscience" (San Antonio, TX, Feb 4-8, 2019) supported by the NSF grants DBI-1820631 (HGR) and IOS-1516648 (FS). The authors acknowledge support from the grants NSF CRCNS-DMS-1608077 (HGR) and NIMH-NIBIB BRAIN Theories1R01EB026939 (FS).

The authors further acknowledge the University of Texas at San Antonio (UTSA) Neuroscience Institute and the New Jersey Institute of Technology (NJIT) Department of Biological Sciences and Institute for Brain and Neuroscience Research for technical support in the organization of the workshop, as well as all of the participants in the workshop.

## List of participants

**Faculty and senior personnel:** Veronica Alvarez, NIH/NIAAA; Asohan Amarsingham, CUNY; William Bechtel, UC San Diego; Carmen Canavier, LSU-HSC; Maurice Chacron, McGill U.; Alain Destexhe, CNRS; Tatiana Engel, CSHL; Hal Greenwald, AFOSR; Andrea Hasenstaub, UC San Francisco; Biyu J He, NYU; Mary-Ann Horn, Case Western Reserve U. ; Kresimir Josic, U. of Houston; Rob Kass, CMU; Aurel Lazar, Columbia U.; William Lytton, SUNY Downstate; Farzan Nadim, NJIT/Rutgers;

Adrien Peyrache, McGill U.; Xaq Pitkow, Rice U.; Astrid Prinz, Emory U.; David Redish, U. of Minnesota; Horacio G Rotstein, NJIT/Rutgers; Fidel Santamaria, UT San Antonio; Sridevi Sarma, Johns Hopkins U.; John White, Boston U; John Rinzel, NYU; Peter Thomas, Case Western Reserve U.; Todd Troyer, UT San Antonio; Zhe Chen , NYU; Renaud Jolivet, CERN/U. of Geneva; Taillefumier Thibaud, UT Austin; Roger Traub, IBM. **Postdocs and research staff:** Enrico Amico, Purdue U.; Aine Byrne, NYU; Bolun Chen , Brandeis U.; Rick Gerkin, Arizona State U.; Sara Ibanez, Marshall U.; Thibault Lagache, Columbia U.; Joseph Monaco, Johns Hopkins U.; Kevin Schultz, Johns Hopkins U.; David Stockton, Emory U.; Ramakrishnan Iyer, Allen Brain Inst.; Yunliang Zang, Okinawa Inst. of Sci. and Tech.; Yiyin Zhou, Columbia U.. **Graduate students**: Habiba Azab, U. of Minnesota; Taku Ito, Rutgers U.; Bhaskar Sen, U. of Minnesota; Sabyasachi Shivkumar, U. of Rochester; Mathew Singh, Washington U.; Dan Levenstein, NYU; Sarah Marzen, MIT; Pake Melland, U. of Iowa;